\documentclass[11pt,superscriptaddress,aps,prd,preprint]{revtex4}
\usepackage{amsmath}

\makeatletter
\usepackage[T1]{fontenc}
\usepackage{amsmath}
\usepackage{amssymb}
\usepackage{graphicx}

\usepackage{slashed}

\usepackage{xcolor}

\newcommand{\bea}{\begin{eqnarray}}
\newcommand{\eea}{\end{eqnarray}}

\begin{document}

\title{On causality violation in Lyra Geometry}

\author{W. D. R. Jesus}\email[]{willian.xb@gmail.com}
\affiliation{Instituto de F\'{\i}sica, Universidade Federal de Mato Grosso,\\
78060-900, Cuiab\'{a}, Mato Grosso, Brazil}

\author{A. F. Santos}\email[]{alesandroferreira@fisica.ufmt.br}
\affiliation{Instituto de F\'{\i}sica, Universidade Federal de Mato Grosso,\\
78060-900, Cuiab\'{a}, Mato Grosso, Brazil}

\begin{abstract}

In this paper the causality issues are discussed in a non-Riemannian geometry, called Lyra geometry. It is a non-Riemannian geometry originated from Weyl geometry. In order to compare this geometry with the Riemannian geometry, the Einstein field equations are considered. It is verified that the G\"{o}del and G\"{o}del-type metric are consistent with this non-Riemannian geometry. A non-trivial solution for G\"{o}del universe in the absence of matter sources is determined without analogue in general relativity. Different sources are considered and then different conditions for causal and non-causal solutions are discussed.
\end{abstract}

 
\maketitle

\section{Introduction}

General Relativity (GR) \cite{Iorio} is a relativistic theory of gravity that accurately explains several gravitational phenomena \cite{Will2006a}. However, it faces some challenges of both observational and theoretical nature (see, e.g., the recent overview \cite{Debono} and references therein), such as accelerated expansion of the universe and the presence of a mysterious form of matter which cannot be directly observed. In addition, there is still no complete and consistent quantum theory of gravity.  From these facts emerges the possibility that GR may be modified. Several modified gravity theories have been developed, such as F(R) gravity \cite{DeFelice2010, Ivan}, bumblebee model \cite{Bluhm2005b}, Chern-Simons gravity \cite {Jackiw2003a}, Brans-Dicke theory \cite{Brans1961}, F(R,T) gravity \cite{Harko}, higher derivative gravity \cite{Hassan}, hybrid metric-Palatini gravity \cite{Olmo}, conformally coupled general relativity \cite{Boris}, bigravity \cite{Vladimir}, non-commutative space-times \cite{Eckstein}, de Sitter Horndeski models \cite{Nunes}, gravity with Lorentz violation \cite{Bailey}, orbital effects of Lorentz-violating gravitomagnetism \cite{Iorio2}, Chameleonic theories \cite{Zanzi}, generalized f(R,$\Phi$, X) gravity \cite{Lobo}, supergravity \cite{Yoshi}, arctan-gravity \cite{Kru}, f(T) gravity \cite{Iorio3, Iorio4}, among others.

In 1918, H. Weyl \cite{Weyl1918, Weyl1919} presented a unified field theory, where the aim was to unify gravitation and electromagnetism. A recent revisitation of Weyl ideas has been done \cite{Barcelo}. In this theory a gauge function that geometrizes the electromagnetism has been introduced. Then the geometry becomes non-Riemannian. The metric compatibility conditions are changed, which means that the related connection is no longer represented by the Christoffell symbol. The major problem with this theory is that the vector length under parallel transport is not constant. This implies the theory is inconsistent, since it becomes non-integrable.

G. Lyra, in 1951, proposed a new modification of the Riemannian geometry \cite{Lyra1951}. Its modification consists in the introduction of a gauge function in the affine space. The gauge function acquires the same status of coordinate system, and together they form the so-called "reference system". Lyra theory is considered as a generalization of the Weyl geometry, but as an advantage the vector length under parallel transport remains constant. This theory has been developed by various authors. For example, Sen \cite{Sen1957} formulated a generalization of GR in the context of Lyra geometry where the gauge function naturally gives rise a displacement vector in the field equations. Halford \cite{Halford1970} has shown that the constant displacement vector plays the role of cosmological constants, with the advantage that the displacement vector arises naturally from the introduction of the gauge function, whereas in GR the cosmological constant is introduced in an ad hoc manner. Sen and Dunn \cite{Sen1971} have shown that the field equations obtained in \cite{Sen1957} can be considered as a special case of the Brans-Dicke theory \cite {Brans1961} if the gauge function is identified as the scalar field.  Beesham \cite{Bee} has studied FLRW cosmological models in Lyra manifold with time dependent displacement field. Recently, numerous studies have been done in the Lyra context, such as, an inhomogeneous Bianchi-I model in Lyra geometry has been studied \cite{Bera}, a thermodynamic analysis of gravitational field equations has been investigated \cite{Moradpour2017}, dark energy has been discussed \cite{Singh2017}, Bianchi type VI cosmological model in the presence of electromagnetic fields with variable magnetic permeability has been analyzed \cite{Abdel-Megied2016},  bulk viscous with strange quark matter attached to the string cloud for higher dimensional Friedman-Robertson-Walker (FRW) universe has been studied \cite{Caglar2016}, static traversable wormholes in Lyra manifold have been considered \cite{Moradpour}, among others. In this paper the main objective is studying the G\"{o}del-type solutions in Lyra geometry looking for a new interpretation for the displacement vector and also solutions that avoid the causality violation.

The G\"{o}del solution is an exact solution of GR proposed by K. G\"{o}del in 1949 \cite{Godel49}. It is the first cosmological solution with rotating matter. This solution represents a universe model that exhibit causality violation since allows the existence of Closed Time-like Curves (CTC's). There are some others models that display CTC's, such as the van Stockum space-time of a rotating dust cylinder \cite{Van}, the Kerr space-time \cite{Kerr} and the Gott space-time of two cosmic strings \cite{Gott}. A generalized version of G\"{o}del metric has been developed which is called G\"{o}del-type metric \cite{PRD28}. In this generalization the causality problem was examined with more details. Here the causality problem is investigated in Lyra geometry where new causal solutions are found.

This paper is organized as follows. In section II, a brief introduction to Lyra geometry is presented. In section III, the G\"{o}del and G\"{o}del-type universes are introduced. In section IV, the G\"{o}del universe in Lyra geometry is studied. In section V, the G\"{o}del-type universe in Lyra geometry is analyzed. The cases $\Lambda=0$ and $\Lambda\neq 0$ are considered for two different energy-momentum tensors. In section VI, some concluding remarks are made.

\section{Lyra Geometry}

Here a brief introduction to Lyra geometry \cite{Lyra1951} is presented. Lyra proposed a generalization of Riemann geometry. In this formalism the components of the displacement vector between two points $P(x^\mu)$ and $P'(x^\mu +dx^\mu)$ are given as $\xi^\mu=x^0dx^\mu$, where $x^0$ is a gauge function. The coordinates system $x^\mu$ and the gauge function $x^0$ form the reference system, i.e, $(x^0;x^\mu)$. Note that, $x^0$ has nothing to do with the time component of the
space-time coordinates. It is a symbol that represents the gauge function. 

The connection components in Lyra geometry are obtained by transforming the general coordinates of the reference system. The coordinate transformation of a multi-component tensor is given by \cite{Sen1957}
\begin{equation}
\xi^{\rho_{1'}\cdots \rho_{s'}}_{\sigma_{1'}\cdots\sigma{r'}}=\lambda^{s-r}A^{\rho_{1'}}_{\rho_1}\cdots A^{\rho_{s'}}_{\rho_s}A^{\sigma_1}_{\sigma_{1'}}\cdots A^{\sigma_r}_{\sigma_{r'}}\xi^{\rho_{1}\cdots \rho_{s}}_{\sigma_{1}\cdots\sigma{r}},
\end{equation}
where $A^{\rho '}_{\rho}\equiv \frac{\partial x^{\rho'}}{\partial x^{\rho}}$ and $\lambda=\frac{x^{0'}}{x^0}$. The factor $\lambda^{s-r}$ arises as a consequence of the introduction of the gauge.
Assuming that a vector $\xi^\mu$ $(x^0;x^\mu)$  is constant in a reference system, such that $\xi^\mu{}_{,\lambda}=0$. 
Then, in another reference system $(x^{0'};x^{\mu '})$ we get
\begin{equation}
\frac{1}{x^0}\xi^{\mu '}{}_{,\lambda '}-\Gamma^{\mu '}{}_{\nu ' \lambda '}\xi^{\nu '}-\frac{1}{2}\phi_{\lambda '}\xi^{\mu '}=0,
\end{equation}
with
\begin{equation}
\Gamma^{\mu '}{}_{\nu ' \lambda '}=-\frac{1}{x^0}A^\alpha_{\nu '}A^{\mu '}_\alpha{}_{\lambda '},\quad\quad\quad \mathrm{and} \quad\quad\quad \phi_{\lambda '}=\frac{1}{x^0}\frac{\partial (\ln \lambda^2)}{\partial x^{\lambda '}}.
\end{equation}

The parallel transport of a vector $\xi^\mu$ in Lyra geometry is given by
\begin{equation}\label{eq:PTL}
d\xi ^\mu=-\tilde{\Gamma}^\mu{}_{\alpha \beta}\xi^\alpha x^0 dx^\beta,
\end{equation}
where
\begin{equation}\label{eq:CNXL}
\tilde{\Gamma}^\mu{}_{\alpha \beta}=\overline{\Gamma}^\mu{}_{\alpha \beta}-\frac{1}{2}\delta^\mu{}_\alpha\phi_\beta,
\end{equation}
with $\tilde{\Gamma}^\mu{}_{\alpha \beta}$ being not symmetric, although $\overline{\Gamma}^\mu{}_{\alpha \beta}$ is symmetric, which the explicit form will be determined.

The parallel transport in Lyra geometry is integrable. The length of the displacement vector is given by the invariant
\begin{equation}
ds^2=g_{\mu \nu}x^0dx^\mu x^0dx^\nu=g_{\mu \nu}\xi ^\mu \xi^\nu.
\end{equation}
Then  $\delta (g_{\mu \nu}\xi ^\mu \xi^\nu)=0$, i.e., the length of a vector is conserved upon parallel transport, as in Riemannian geometry. Using this condition, we get
\begin{equation}\label{eq:CNXW}
\overline{\Gamma}^\mu{}_{\alpha \beta}=\frac{1}{x^0}\Gamma^\mu{}_{\alpha \beta}+\frac{1}{2}(\delta^\mu{}_\alpha\phi_\beta+\delta^\mu{}_\beta\phi_\alpha-g_{\alpha \beta}\phi^\mu),
\end{equation}
that disregarding the factor $(1/x^0)$ is identical to the connection obtained by Weyl \cite{Weyl1919}. Note that, $\Gamma^\mu{}_{\alpha \beta}$ is the Christoffel symbol in Riemannian geometry.

The curvature tensor in Lyra geometry is given as
\begin{equation}
K^\mu{}_{\lambda \alpha \beta}=(x^0)^{-2}\left[\frac{\partial(x^0\tilde{\Gamma}^\mu{}_{\lambda \beta})}{\partial x^\alpha}-\frac{\partial(x^0\tilde{\Gamma}^\mu{}_{\lambda \alpha})}{\partial x^\beta}+x^0\tilde{\Gamma}^\mu{}_{\rho \alpha}x^0\tilde{\Gamma}^\rho_{\lambda \beta}-x^0\tilde{\Gamma}^\mu{}_{\rho \beta}x^0\tilde{\Gamma}^\rho_{\lambda \alpha}\right].
\end{equation}
It is obtained in the same as in general relativity. From it the Ricci tensor is
\bea
K_{\kappa\alpha}\equiv K^\mu{}_{\kappa \mu \alpha}=\frac{R_{\kappa\alpha}}{(x^0)^2}+\frac{1}{x^0}\phi_{\kappa;\alpha}+\frac{1}{2x^0}g_{\kappa\alpha}\phi^i\,_{;i}-\frac{1}{2}g_{\kappa\alpha}\phi^\gamma\phi_\gamma+\frac{1}{2}g_{\kappa\alpha}\phi^\circ_i\phi^i
\eea
with $R_{\kappa\alpha}$ being the Ricci tensor of Riemann geometry and $\phi^\circ_\alpha\equiv(x^0)^{-1}\frac{\partial[ln (x^0)^2]}{\partial x^\alpha}$. The curvature scalar is
\begin{equation}
K\equiv g^{\kappa\alpha}K_{\kappa\alpha}=(x^0)^{-2}R+3(x^0)^{-1}\phi^\alpha{}_{;\alpha}+\frac{3}{2}\phi_\alpha\phi^\alpha+\frac{3}{2}\phi^\circ_\alpha\phi^\alpha,
\end{equation}
where $R$ is the curvature scalar in the Riemannian geometry. If we choose the normal gauge, i.e., $x^0=1$, the curvature scalar becomes
\begin{equation}
K=R+3\phi^\alpha{}_{;\alpha}+\frac{3}{2}\phi_\alpha\phi^\alpha.
\end{equation}

The field equations are obtained from the variational principle
\begin{equation}
\delta\int\sqrt{-g}(K+\mathcal{L}-2\Lambda)x^0 dx^1 \cdots x^0dx^4=0,
\end{equation}
where $\mathcal{L}$ is the matter Lagrangian density and $\Lambda$ is the cosmological constant. Considering $\phi_\alpha$ as an external field, the field equations are obtained varying the action with respect to $g_{\alpha \beta}$. Then the Einstein equations in Lyra geometry are
\begin{equation}\label{eq:FE1}
R_{\mu \nu}-\frac{1}{2}Rg_{\mu \nu}+\Lambda g_{\mu\nu}+\frac{3}{2}\phi_\mu \phi_\nu-\frac{3}{4}g_{\mu \nu}\phi_\alpha\phi^\alpha=8\pi GT_{\mu \nu},
\end{equation}
where $T_{\mu \nu}$ is the energy-momentum tensor associated with the matter distribution. Multiplying eq. (\ref{eq:FE1}) by  $g^{\mu \nu}$, the field equations are rewritten as
\begin{equation}\label{eq:FE2}
R_{\mu \nu}=8\pi G(T_{\mu \nu}-\frac{1}{2}Tg_{\mu \nu})-(L_{\mu \nu}-\frac{1}{2}Lg_{\mu \nu})-\Lambda g_{\mu\nu},
\end{equation}
where $T=g^{\mu \nu}T_{\mu \nu}$ is the trace of the energy-momentum tensor, $L_{\mu \nu}$ is defined as
\begin{equation}
L_{\mu \nu}=\frac{3}{2}\phi_\mu \phi_\nu-\frac{3}{4}g_{\mu \nu}\phi_\alpha\phi^\alpha,
\end{equation}
and $L$ is its trace, i.e., $L=g^{\mu \nu}L_{\mu \nu}$.

\section{G\"{o}del and G\"{o}del-Type Universes}

The G\"{o}del metric \cite{Godel49} is the first exact solution with rotating matter of Einstein field equations. This solution is stationary, spatially homogeneous and has cylindrical symmetry. The main characteristic of G\"{o}del metric is the possibility of closed timelike curves (CTC's), that imply causality violation, i.e., in theory travel to the past is allowed. The G\"{o}del metric is
\begin{equation}\label{eq:11}
ds^2=a^2\left[dt^2-dx^2\frac{1}{2}e^{2x}dy^2-dz^2+2e^xdtdy\right],
\end{equation}
where $a$ is a positive constant. The Ricci tensor and Ricci scalar associated with this metric are
\begin{equation}\label{eq:12}
R_{\mu \nu}=\frac{1}{a^2} u_\mu u_\nu \qquad \mathrm{and} \qquad R=\frac{1}{a^2}.
\end{equation}
Considering the energy-momentum tensor as $T_{\mu \nu}=\rho u_\mu u_\nu$, where $\rho$ is the matter density and $u_\mu=(a,0,ae^x,0)$ is the 4-velocity, the Einstein equations, 
\begin{equation}\label{eq:13}
R_{\mu \nu}-\frac{1}{2}Rg_{\mu \nu}=8\pi GT_{\mu \nu}+\Lambda g_{\mu \nu},
\end{equation}
are satisfied for the conditions
\begin{equation}\label{eq:14}
8\pi G\rho=\frac{1}{a^2} \qquad \mathrm{an}d \qquad \Lambda=-\frac{1}{2a^2}.
\end{equation}

The G\"{o}del solution may be considered as a family member of homogeneous geometries of  G\"{o}del-type space-time \cite{PRD28}. In cylindrical coordinates  G\"{o}del-type metric is
\begin{equation}\label{eq:15}
ds^2=[dt+H(r)d\phi]^2-dr^2-D^2(r)d\phi^2-dz^2,
\end{equation}
where the functions $H(r)$e $D(r)$ are such that
\begin{equation}\label{eq:16}
\frac{H'}{D}=2\omega \qquad and \qquad \frac{D''}{D}=m^2,
\end{equation}
where the prime means derivative with respect to $r$. The parameters $m$ and $\omega$ are constants used to classify different G\"{o}del-type geometries, such that $\omega >0$ and $-\infty\leq m^2 \geq +\infty$.
The metric, eq. \eqref{eq:15}, is rewritten as
\begin{equation}\label{eq:17}
ds^2=dt^2+2H(r)dtd\phi-dr^2-G(r)d\phi^2-dz^2,
\end{equation}
where $G(r)=D^2(r)-H^2(r)$. In this form, the existence of CTC is evident and the circles defined by $t,z,r=constant$ depend on the behavior of $G(r)$. If $G(r)<0$ for a certain range of $r$, the G\"{o}del circles are CTC's. The causal characteristics of G\"{o}del-type space-times depend on the two independent parameters $m$ and $\omega$. The G\"{o}del-type metric is classified into three different classes:

\begin{itemize}
\item[i)] Linear ($m=0$) with
\begin{equation}\label{eq:18}
H(r)=\omega r^2 \qquad \mathrm{and} \qquad D(r)=r.
\end{equation}
The critical radius is $r_c=\frac{1}{\omega}$, such that for all $r>r_c$ there are non-causal G\"{o}del circles. 
\item[ii)] Trigonometric ($m^2 \equiv \mu ^2<0$) with
\begin{equation}\label{eq:19}
H(r)=\frac{4\omega}{\mu^2}sin^2\left(\frac{\mu r}{2}\right) \qquad \mathrm{and} \qquad D(r)=\frac{1}{\mu}sin(\mu r).
\end{equation}
Here exist an infinite sequence, alternating between causal and non-causal regions.
\item[iii)] Hiperbolic ($m^2>0$) with
\begin{equation}\label{eq:110}
H(r)=\frac{4\omega}{m^2}sinh^2\left(\frac{mr}{2}\right) \qquad \mathrm{and} \qquad D(r)=\frac{1}{m}sinh(mr).
\end{equation}
\end{itemize}
The hyperbolic class may be separated into two, depending on $m^2$: (1) $0< m^2<4\omega^2$, where there is one non-causal region for $r>r_{c}$, with the critical radius $r_c$ given by
\begin{equation}\label{eq:111}
sinh^2\left(\frac{mr_c}{2}\right)=\left(\frac{4\omega^2}{m^2}-1\right)^{-1}.
\end{equation}
(2) $m^2\geq 4\omega^2$, in this case there is no breakdown of causality and, thus, no occurrence of CTCs,  i.e.,  $r_c \rightarrow \infty$ (completely causal region). If $m^2=2\omega^2$, the  G\"{o}del result is recovered \cite{Godel49}.

In order to study the field equations, for the sake of simplicity, let's use the following basis \cite{PRD28}
\begin{equation}\label{eq:114}
\theta^0=dt+H(r)d\phi, \qquad \theta^1=dr, \qquad \theta^2=D(r)d\phi, \qquad \theta^3=dz,
\end{equation}
where the line element takes the form
\begin{equation}\label{eq:113}
ds^2=\eta_{AB}\theta ^A\theta ^B=(\theta^0)^2-(\theta^1)^2-(\theta^2)^2-(\theta^3)^2,
\end{equation}
with $\theta^A=e^A{}_\mu dx^\mu$ being the 1-forms and $\eta_{AB}=diag(+1,-1,-1,-1)$ being the Minkowski metric. In this new basis, a flat space-time, the Ricci tensor is given by
\begin{equation}\label{eq:115}
R_{00}=\frac{1}{2}\left(\frac{H'}{D}\right)^2=2\omega ^2, \qquad R_{11}=R_{22}=R_{00}-\frac{D''}{D}=2\omega^2-m^2,
\end{equation}
and the scalar curvature is
\begin{equation}\label{eq:116}
R=2(m^2-\omega^2).
\end{equation}
Then Einstein tensors become
\begin{equation}\label{eq:117}
G_{00}=3\omega^2-m^2, \qquad G_{11}=G_{22}=\omega^2, \qquad G_{33}=m^2-\omega^2.
\end{equation}

\section{G\"{o}del Solution in Lyra Geometry}

In this section the energy-momentum tensor is considered as
\begin{equation}
T_{\mu \nu}=\rho u_\mu u_\nu,
\end{equation}
with  $u_\mu=(a,0,ae^x,0)$ and its trace as $T=\rho$. In order to calculate the field equations, eq. (\ref{eq:FE2}), the displacement vector is chosen as
\begin{equation}\label{eq:VD02g}
\phi_\alpha=(\frac{2}{\sqrt{3}}ab,0,\frac{2}{\sqrt{3}}abe^x,0),
\end{equation}
where $b$ is a constant. Using G\"{o}del metric, eq. (\ref{eq:11}), the field equations become
\begin{align}
\frac{2}{a^2}&=\kappa \rho-2\Lambda-4b^2,\label{eq:VD02g a}\\
\frac{4}{a^2}&=3\kappa \rho-2\Lambda-8b^2,\label{eq:VD02g b}\\
\kappa \rho&=-2\Lambda\label{eq:VD02g c}.
\end{align}
Rearranging equations \eqref{eq:VD02g a} and \eqref{eq:VD02g c} we obtain
\bea
\kappa \rho&=&\frac{1}{a^2}+2b^2,\\
\Lambda&=&-\frac{1}{2a^2}-b^2.
\eea

This result modifies the standard G\"{o}del relations, although it is an exact solution of field equations. Then CTC's exist in Lyra geometry for the displacement vector~\eqref{eq:VD02g}, i.e., there is a causality violation in this theory. An important note, for the case that $\Lambda=0$, the solution is $b^2=-\frac{1}{2a^2}$ and $\kappa\rho=0$. Therefore, in Lyra geometry for an empty universe the $b^2$ parameter is equal to the cosmological constant in Riemann geometry. Similar interpretations involving $b^2$ and $\Lambda$ have been presented \cite{Sen1957, Singh1993, Halford1970}. Other choices for $\phi_\alpha$ lead to solutions with $b^2=0$, i.e., there is no contribution due to Lyra geometry.

\section{G\"{o}del-Type Solution in Lyra Geometry}

In order to study the G\"{o}del-type solution in Lyra geometry two cases are considered: (i) zero cosmological constant and (ii) non-zero cosmological constant. 

\subsection{$\Lambda=0$} 

Here the field equations, eq. \eqref{eq:FE2}, in the tetrad frame become
\begin{equation}
R_{AB}=\kappa(T_{AB}-\frac{1}{2}T\eta_{AB})-(L_{AB}-\frac{1}{2}L\eta_{AB}).
\end{equation}
To solve this equation, let's consider as source of matter a perfect fluid. Its energy-momentum tensor is
\begin{equation}\label{eq:EnergyM Tensor}
T_{AB}=(\rho+p)u_Au_B-p\eta_{AB},
\end{equation}
where $\rho$ is the energy density, $p$ is the pressure and $u_A=(1,0,0,0)$ is the 4-velocity of the fluid. The trace of eq. \eqref{eq:EnergyM Tensor} is 
\begin{equation}
T=\eta^{AB}T_{AB}=\rho-3p.
\end{equation}

Considering the displacement vector as
\begin{equation}\label{VecDes b000}
\phi_A=\left(\frac{2}{\sqrt{3}}b,0,0,0 \right),
\end{equation}
the field equations become
\begin{align}
2\omega^2+2b^2 &=\frac{1}{2}\kappa(\rho+3p) \label{Field Equations b000a},\\
2\omega^2-m^2 &=\frac{1}{2}\kappa(\rho-p)\label{Field Equations b000b},\\
\rho &=p.\label{Field Equations b000c}
\end{align}
From these equations we get
\begin{align}
&m^2=2\omega^2,\label{Field Equations b000d}\\
&\omega^2+b^2=\kappa \rho.\label{Field Equations b000e}
\end{align}

The condition \eqref{Field Equations b000d} corresponds to G\"{o}del solution that leads to causality violation. As consequence of the new geometry the critical radius is modified and becomes
\bea
r_c=\frac{2}{m}\sinh^{-1}(1)=2\sinh^{-1}(1)\frac{1}{\sqrt{2\kappa\rho-2b^2}}.
\eea
 In Riemann geometry, the parameters are related as $m^2=2\omega^2=\kappa \rho=-2\Lambda$. Here the same condition is obtained from eq. \eqref{Field Equations b000e} with
\begin{equation}
b^2=\omega^2=-\Lambda,
\end{equation}
i.e., $b^2$ corresponds to the negative cosmological constant. Then the displacement vector has a complex value. A similar result has been found in \cite{Sen1957}.

As a second choice, let's consider
\begin{equation}
\phi_A=\left(0,0,0,\frac{2}{\sqrt{3}}b\right)\label{VDtg 3},
\end{equation}
which leads to the field equations 
\begin{align}
2\omega^2 &=\frac{1}{2}\kappa(\rho+3p),\label{eq: 000ba}\\
2\omega^2-m^2&=\frac{1}{2}\kappa(\rho-p),\label{eq: 000bb}\\
2b^2 &=\frac{1}{2}\kappa(\rho-p)\label{eq: 000bc}.
\end{align}

Combining these equations we obtain
\begin{equation}\label{eq: 000bd}
m^2=2\omega^2-2b^2. 
\end{equation}
Here, two possibles class of solution may be considered. The first class corresponds to the Som-Raychaudhuri space-time \cite{Som1968}. In this case $m^2=0$ and then $b^2=\omega^2=constant$. The second class corresponds to $m^2>0$. Here there are two different cases: (i) Non-causal: If $b^2=0$, the G\"{o}del solution with $m^2=2\omega^2$ is recovered; (ii) Causal case: If $b^2=-\omega^2$, the causal relation, $m^2=4\omega^2$, is obtained. Another attempt to get a causal solution is to consider $\rho=-p$, the exotic distribution of matter, that leads to the causal solution $m^2=4\omega^2$. This last result leads to a complex value for vorticity $\omega$. Other choices for $\phi_A$ imply solutions with $b^2=0$. 

Another way to obtain a causal solution is to consider the energy-momentum tensor as
\begin{equation}
T_{AB}=T^{M}_{AB}+T^{S}_{AB}\label{eq:EnergyMScalar},
\end{equation}
where $T^{M}_{AB}$ is the energy-momentum tensor of the perfect fluid and $T^S_{AB}$ is the energy-momentum tensor associated to the scalar field $\Phi$, given by 
\begin{equation}
T^{S}_{AB}=\nabla_A\Phi\nabla_B\Phi-\frac{1}{2}\eta_{AB}\eta^{CD}\nabla_C\Phi\nabla_D\Phi,
\end{equation}
with $\nabla_A$ being the covariant derivative with respect to base $\theta^A=e^A{}_Bdx^B$.  Assuming, for simplicity, that
\begin{equation}
\Phi(z)=ez+constant,
\end{equation}
the energy-momentum tensor components associated with the scalar field are
\begin{equation}
T^S_{00}=-T^S_{11}=-T^S_{22}=T^S_{33}=\frac{e^2}{2}
\end{equation}
and its trace is $T=\rho-3p+e^2$. Let's analyze the cases where the $\phi_A$ implies no-trivial solutions, i.e., eqs. \eqref{VecDes b000} and \eqref{VDtg 3}. For the choice \eqref{VecDes b000}, the field equations become
\begin{align}
2\omega^2+2b^2 &=\frac{1}{2}\kappa(\rho+3p)\label{eq:EQ VD 0 S a},\\
2\omega^2-m^2 &=\frac{1}{2}\kappa(\rho-p),\label{eq:EQ VD 0 S b}\\
-\kappa e^2 &=\frac{1}{2}\kappa(\rho-p)\label{eq:EQ VD 0 S c}.
\end{align}
Combining these equations we get
\begin{equation}
m^2-2\omega^2=\kappa e^2.
\end{equation}
Due to the introduction of the scalar field, there is a freedom to choose the $m^2$ value. Considering that  G\"{o}del-type metric belongs to the hyperbolic class, where $m^2>0$. The possible solutions are: G\"{o}del solution and a causal solution. If $m^2=2\omega^2$,  the G\"{o}del solution is recovered with $e=0$. Note that, $\kappa e^2>0$, then $m^2=4\omega^2$ is the possible value which leads to a causal solution. In this case, the scalar field $e$ assume the value $e=\sqrt{\frac{2}{k}}\omega$.

Now, for the choice \eqref{VDtg 3}, the field equations are
\begin{align}
2\omega^2 &=\frac{1}{2}\kappa(\rho+3p),\\
2\omega^2-m^2&=\frac{1}{2}\kappa(\rho-p),\\
2b^2 -ke^2&=\frac{1}{2}\kappa(\rho-p).
\end{align}
From these equations we obtain
\begin{align}
m^2=2\omega^2-2b^2+\kappa e^2.
\end{align}
For the hyperbolic class we get a causal solution if
\begin{equation}
b^2=\frac{1}{2}\kappa e^2-\omega^2.
\end{equation}
Therefore, in both cases with scalar field the causal solution is possible.

\subsection{$\Lambda\neq0$} 

The field equations for choice \eqref{VecDes b000} become
\begin{align}
2\omega^2+2b^2 &=\frac{1}{2}\kappa(\rho+3p)-\Lambda,\label{FE VDtg 0a}\\
2\omega^2-m^2 &=\frac{1}{2}\kappa(\rho-p)+\Lambda,\label{FE VDtg 0b}\\
-\Lambda&=\frac{1}{2}\kappa(\rho-p)\label{FE VDtg 0c}.
\end{align}
Combining these equations, we get
\begin{align}
m^2 &=2\omega^2,\label{FE VDtg 0d}\\
2\omega^2&=\kappa(\rho+p)-2b^2\label{FE VDtg 0e}.
\end{align}
The equation \eqref{FE VDtg 0d} leads to the G\"{o}del solution. In addition the critical radius is modified, i.e.,
\begin{equation}
r_c=\frac{2}{m}\sinh^{-1}(1)=2\sinh^{-1}(1)\frac{1}{\sqrt{\kappa(\rho+p)-2b^2}}.
\end{equation}
An important note, beyond which there exist non-causal G\"{o}del circles, which depend on the density matter, pressure and the $b^2$ parameter, that is due to Lyra geometry. Therefore, is possible to find a completely causal G\"{o}del solution in Lyra geometry. It occurs for $2b^2=\kappa(\rho+p)$, since $r_c \rightarrow \infty$.

For the displacement vector given by eq. \eqref{VDtg 3}, the field equations are given as
\begin{align}
2\omega^2 &=\frac{1}{2}\kappa(\rho+3p)-\Lambda,\\
2\omega^2-m^2&=\frac{1}{2}\kappa(\rho-p)+\Lambda,\\
2b^2&=\frac{1}{2}\kappa(\rho-p)+\Lambda.
\end{align}
Rearranging the field equations,
\begin{align}
\kappa \rho &=\omega^2-\Lambda+3b^2,\\
\kappa p &=\omega^2+\Lambda-b^2,\\
m^2 &=2\omega^2-2b^2.
\end{align}
For the class solution with $m^2=0$ we get $\omega^2=b^2$. The G\"{o}del solution happens to $b=0$, as expected. A causal solution, with $m^2=4\omega^2$, leads to a complex value for the vorticity.

In order to obtain a different causal solution the scalar field is considered as in eq. \eqref{eq:EnergyMScalar}. Using eq. \eqref{VecDes b000}, the equations system is
\begin{align}
2\omega^2+2b^2 &=\frac{1}{2}\kappa(\rho+3p)-\Lambda,\\
2\omega^2-m^2 &=\frac{1}{2}\kappa(\rho-p)+\Lambda,\\
-\kappa e^2 &=\frac{1}{2}\kappa(\rho-p)+\Lambda.
\end{align}
From these equations, we obtain
\begin{equation}
m^2=2\omega^2+\kappa e^2.
\end{equation}
In the hyperbolic class of solution, the causal solution is obtained for
$2\omega^2=\kappa e^2$, which leads to $b^2=\frac{1}{2}\kappa(\rho + p)$.

For the displacement vector \eqref{VDtg 3}, the field equations are given as
\begin{align}
\kappa \rho &=\omega^2-\Lambda-\frac{3}{2}\kappa e^2+3b^2,\\
\kappa p &=\omega^2+\Lambda+\frac{1}{2}\kappa e^2-b^2,\\
m^2 &=2\omega^2+\kappa e^2-2b^2.
\end{align}
Here, a causal solution emerges  only if $b^2=\frac{1}{2}\kappa e^2 -\omega^2$.

An important note, the universe rotation problem is one of the most intriguing in modern cosmology. There are two opposing points of view on this problem. (i) The universe as a whole is usually considered to be non-rotational in contrast to astronomical objects, such as planets, stars, and galaxies. (ii) The possibility that the universe rotates should not be ignored since the solutions of general relativity corresponding to a rotating universe have been found directly indicating that a global rotation is physically allowed. Although it is an important problem, our results do not address a solution. However, the objective here was to study the causality issue in a solution of general relativity that describes a rotating universe within a non-Riemannian geometry.

\section{Conclusions}

The GR has been developed in a Riemannian geometry. However is possible study gravitational effects in others geometries. Lyra geometry is a non-Riemannian geometry that allows study Einstein equations. Although various studies and applications have been realized in Lyra geometry, there is nothing about causality violation in this context. G\"{o}del solution is an exact solution of GR that exhibits CTC's, then it allows causality violation. In this context is interesting to study this solution in a different geometry. Our analysis is developed for G\"{o}del metric and for G\"{o}del-type metric for physically well-motivated matter sources presented by perfect fluid and a scalar field. Our results have shown that the field equations solution depend on the choice of the displacement vector $\phi_\alpha$. For the choice given by eq. (\ref{eq:VD02g}) the G\"{o}del metric is a solution of Einstein equations in the Lyra manifold. Here the usual GR solution is changed by a constant factor. In addition, for an empty universe, the $b^2$ parameter is equal to the cosmological constant in Riemann geometry. In order to study the G\"{o}del-type solution different cases are considered.\\
1. Perfect fluid as matter source:\\
(i) Displacement vector given by eq. (\ref{VecDes b000}); For $\Lambda=0$ the original G\"{o}del solution is obtained from the condition $m^2=2\omega^2$ that naturally emerges. From this condition the $b^2$ parameter may be interpreted as a negative cosmological constant. For $\Lambda\neq 0$ G\"{o}del solution is obtained again. A critical radius for causality violation is calculated and a condition for causal solution is determined. \\
(ii) Displacement vector given by eq. (\ref{VDtg 3}); With $\Lambda=0$ two class of solution are considered. The first class corresponds to the case $m^2=0$ and then $b^2=\omega^2=constant$. The second class corresponds to $m^2>0$. In this case there are two different cases: (a) Non-causal: If $b^2=0$, G\"{o}del solution with $m^2=2\omega^2$ is recovered; (b) Causal case: If $b^2=-\omega^2$, the causal relation, $m^2=4\omega^2$, is obtained. For $\Lambda\neq 0$ a causal solution is obtained and it leads to a complex value for the vorticity. \\
2. Matter sources - Perfect fluid plus scalar field:\\
(i) Displacement vector given by eq. (\ref{VecDes b000}); $\Lambda=0$ - In this case, the introduction of the scalar field gives freedom to choose the $m^2$ value and then causal and non-causal solutions have been found. When $\Lambda\neq 0$ the causal solution is obtained for $2\omega^2=\kappa e^2$, which leads to $b^2=\frac{1}{2}\kappa(\rho + p)$.\\
(ii) Displacement vector given by eq. (\ref{VDtg 3}); In both cases, $\Lambda=0$ and $\Lambda\neq0$, the causal solution has been determined with a particular value for $b^2$ parameter.  

Therefore, our results shown that in Lyra geometry is possible avoid the causality violation choosing a particular value for a parameter that comes from geometry. In addition, for a perfect fluid as matter source a causal solution is determined. This result allows an analyze that there is no analogue in GR.

\section*{Acknowledgments}

This work by A. F. S. is supported by CNPq project 308611/2017-9.

\end{document}